\documentclass[12pt]{article}
\input epsf.sty
\newcommand{\nl}{\hspace{-.65cm}}
\newcommand{\be}{\begin{equation}}
\newcommand{\ee}{\end{equation}}
\newcommand{\ben}{\begin{eqnarray}\displaystyle}
\newcommand{\een}{\end{eqnarray}}

\usepackage{amssymb}
\usepackage{amssymb}
\usepackage{amsmath}
\usepackage{comment}
\usepackage{mathrsfs}
\usepackage[all]{xy}
\usepackage[dvips]{graphicx}
\usepackage{epsfig,color}

\def\be{\begin{equation}}
\def\ee{\end{equation}}
\def\ba{\begin{align}}
\def\ea{\end{align}}

\begin{document}

\baselineskip=18pt

\begin{center}
{\Large \bf{ Stringy Horizons and UV/IR Mixing}}

\vspace{10mm}


\vspace{10mm}

\textit{Roy Ben-Israel$\,^1$, Amit Giveon$\,^2$, Nissan Itzhaki$\,^1$ and Lior Liram$\,^1$}
\break

$^1$ Physics Department, Tel-Aviv University, Israel	\\
Ramat-Aviv, 69978, Israel \\

$^2$ Racah Institute of Physics, The Hebrew University \\
Jerusalem, 91904, Israel
\break

\end{center}

\begin{abstract}

The target-space interpretation of the exact (in $\alpha'$) reflection coefficient
for scattering from Euclidean black-hole horizons in classical string theory
is studied.
For concreteness, we focus on the solvable
$SL(2, \mathbb{R})_k/U(1)$ black hole.
It is shown that it exhibits a fascinating UV/IR mixing,
dramatically modifying the late-time behavior of general relativity.
We speculate that this might play an important role in the
black-hole information puzzle,
as well as in clarifying features related
with the non-locality of Little String Theory.

\end{abstract}

\newpage

\section{Introduction}

Since the work of Bekestein \cite{Bekenstein:1973ur} and Hawking \cite{Hawking:1974sw}, the study of black holes at the quantum level received much attention. In comparison, classical stringy corrections ($g_s=0$ and finite $\alpha'=l_s^2$) received little attention.
The reason is that  it is widely believed that they are negligible for large black holes. Indeed, it has been known for a while \cite{Callan:1988hs} that  perturbative $\alpha'$
corrections have a tiny  effect on the physics of large black holes. This is well understood, as such corrections are described by irrelevant terms that are parametrically small at the horizon of a large black hole.

Motivated by \cite{Kutasov:2005rr}, we recently argued \cite{Giveon:2012kp,Giveon:2013ica,Giveon:2013hsa,Giveon:2014hfa,I} (for recent related and  potentially related works see \cite{Mertens:2013pza,Mertens:2013zya,Mertens:2014cia,Mertens:2014dia,Mertens:2014saa,Mertens:2015hia,Giribet:2015kca} and \cite{Dodelson:2015toa,Dodelson:2015uoa}, respectively) that this is not the case for non-perturbative  $\alpha'$  corrections, which play an important role at the horizon. The argument  is clearer in the Euclidean version of the black hole -- the cigar geometry. It is particularly sharp for the cigar sigma-model obtained in the
coset  $SL(2, \mathbb{R})_k/U(1)$, since then we have an exact CFT description.

In this paper, we continue the study of strings on $SL(2, \mathbb{R})_k/U(1)$.
Following \cite{I}, we focus our attention on scattering amplitudes in this geometry.
To allow on-shell states in string theory,
we add an extra time direction, $t$, and consider large $t$ processes in the
small curvature (large $k$) limit.
One may suspect that in this case stringy corrections have negligible effects. The main point of the paper is to show that this is not the case, and that non-perturbative  $\alpha'$ corrections in fact lead to an interesting UV/IR mixing that could potentially lead to a better understanding of the black-hole information puzzle and reveal new features of Little String Theory (LST).

The paper is organized as follows. In section 2, we review the set-up and calculation of the reflection coefficient, associated with scattering in the cigar geometry. In section 3, we show that the exact reflection coefficient  \cite{Teschner:1999ug} leads to UV/IR mixing. In section 4, we discuss the physics that led to this UV/IR mixing and, in section 5, we emphasise the sensitivity of the results to coarse graining. Section 6 is devoted to a discussion and, finally, some details are presented
in an appendix.

\section{The reflection coefficient}

In this section, we review the reflection coefficient associated with the coset CFT, $SL(2, \mathbb{R})_k/U(1)$. We begin by describing the setup, and then move on to recall the derivation of the classical reflection coefficient. Finally, we briefly review the exact CFT result.

\subsection{The setup}

We wish to study the coset CFT, $SL(2, \mathbb{R})_k/U(1)$, whose sigma-model background takes the form of the cigar geometry \cite{Elitzur:1991cb,Mandal:1991tz,Witten:1991yr,Dijkgraaf:1991ba},
\be\label{cigar}
ds^2=2k \tanh^2 \left(\frac{\rho}{\sqrt{2 k}}\right) d\theta^2
+d\rho^2~,~~~~
\exp(2 \Phi )= \frac{g_{0}^2}{\cosh^2 \left(\frac{\rho}{\sqrt{2 k}}\right)}~.~~~~
\ee
The angular direction $\theta$ has periodicity $2\pi$, compatible with
smoothness of the background at the tip, and $\Phi$ is the dilaton.
We work with $\alpha'=2$.
In the supersymmetric case,
the background (\ref{cigar}), which is obtained e.g.
by solving the graviton-dilaton e.o.m. in the leading GR approximation,
is perturbatively {\it exact} in $\alpha'$ \cite{Bars:1992sr,Tseytlin:1993my}.

\begin{figure}
\centerline{\includegraphics[scale=0.34]{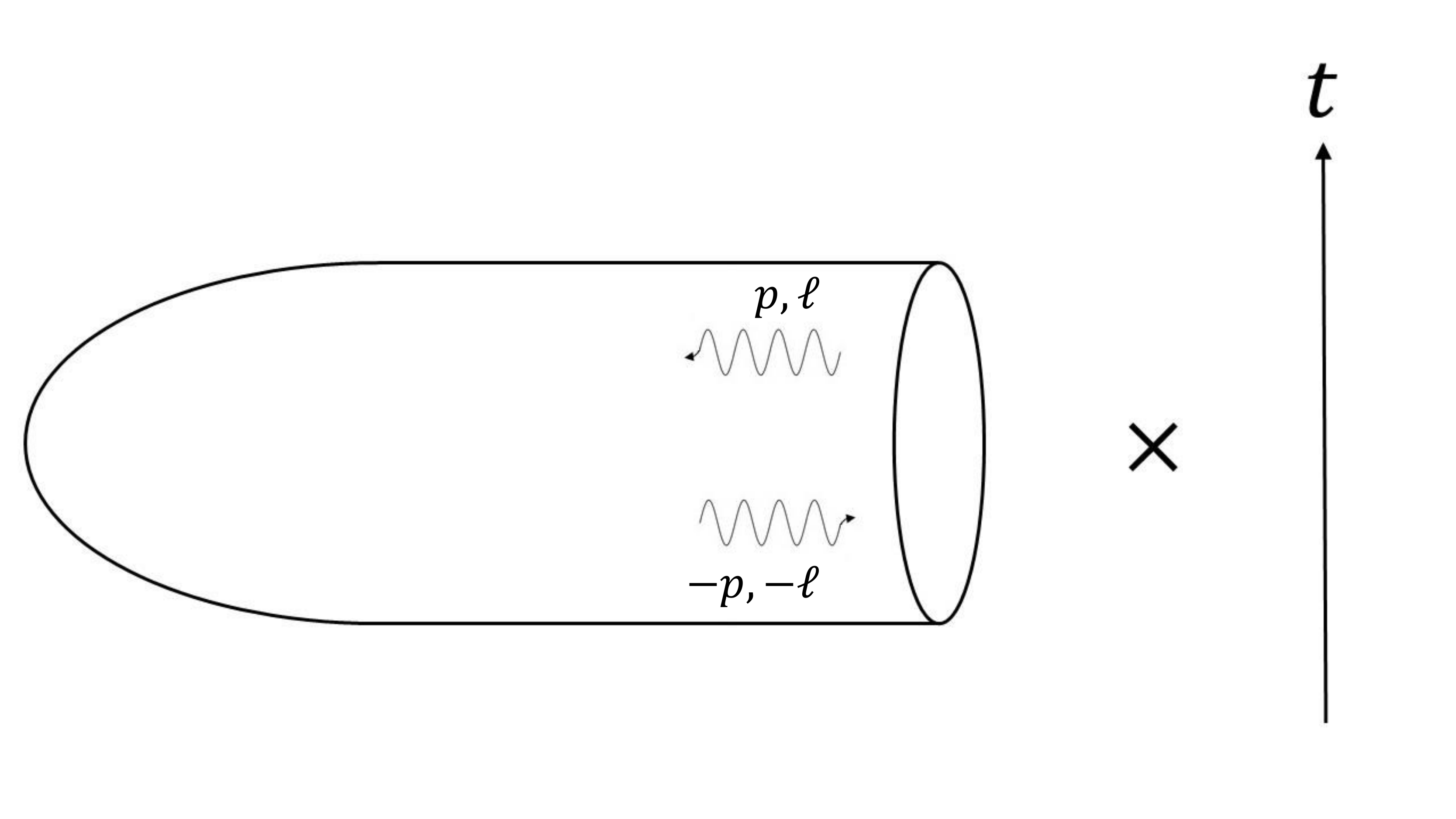}}
\caption{A plot of the classical geometry induced from the $SL(2, \mathbb{R})_k/U(1)$ coset model -- a cigar that pinches off at $\rho=0$, which upon Wick rotation, corresponds to the horizon. Asymptotically, the classical solution of the Klein-Gordon equation is a
sum of incoming and outgoing radial waves with radial momentum $p$ and angular momentum $\ell$. The time direction $t$ imposes the on-shell condition.
The manifold $M$ is not pictured.
}
\label{cigarplot}
\end{figure}

We intend to probe this model with primary fields that correspond to high-energy modes,
which propagate on this background (see figure \ref{cigarplot}). The reflection coefficients associated with such modes are known exactly \cite{Teschner:1999ug}. We shall take advantage of this fact and attempt to extract some target-space information about the cigar theory, which hopefully provides us with data that goes beyond the GR solution (\ref{cigar}).

In terms of the underlying $SL(2, \mathbb{R})$ representation,
these modes are in the continuous representations,
\be\label{js}
j=-1/2+is, ~~s\in \mathbb{R},
\ee
where $s$ is related to the momentum in the radial direction.
More precisely, it is
\be
s=\sqrt{\frac{k}{2}} \, p,
\ee
where $p$ is the momentum associated with the canonically-normalized field $\rho$,
corresponding to the radial direction, at $\rho\to\infty$.
Additional quantum numbers, associated with the probing modes,
are $m$ and $\bar{m}$, which are related to $\ell$,
the angular momentum along the $\theta$ direction,
and $\omega$, the winding number on the semi-infinite cigar, via
\be\label{mmpw}
(m, \bar{m})=\frac12 (\ell+k \omega,-\ell+k \omega).
\ee
Our main focus is on GR-like modes, with $w=0$, so that $m=-\bar{m}=\ell/2$.

We would like these modes to correspond to on-shell states in string theory.
Hence, we add an auxiliary time direction, $t$.
Denoting the energy associated with $t$ by $E$, the on-shell condition reads
\be\label{onshell}
E^2=p^2+\frac{\ell^2+1}{2 k}.
\ee
A compact manifold, $M$, that plays no role here,
is also added so that the string-theory background is critical.

Note that this setup is (a particular, weakly coupled) Little String Theory,
which is closely related to Double-Scaled LST
(see e.g. \cite{Aharony:2004xn} and references therein).
Thus, we are studying here
physics associated with peculiarities, such as {\it non-locallity}, of LST;
we shall discuss this further in section 6.

The exact reflection coefficient can be written as a product of the
perturbative (GR) reflection coefficient and a non-perturbative correction in $\alpha'$
\cite{Teschner:1999ug},
\be\label{ftg}
R(j;m,\bar{m})= R_{pert}(j;m, \bar{m})R_{non-pert}(j).
\ee
The perturbative part, as the name implies,
can be derived from GR plus perturbative $\alpha'$ corrections.
In the supersymmetric case, on which we focus on here,
there are no perturbative $\alpha'$ corrections and, as we review below,
$R_{pert}(j;m, \bar{m})$ can be derived by solving the wave equation in
the curved background (\ref{cigar}).

The non-perturbative part comes from the exact computation of the reflection coefficient
in the quotient CFT.
Its physical origin can be traced to the condensation of a tachyon field,
associated with a winding string near the tip of the cigar.


\subsection{The GR reflection coefficient}

For a scalar field $a(\rho,\theta,t)$, $R_{pert}$ can be obtained  by solving the Klein-Gordon equation on the cigar background (\ref{cigar}). This equation takes the form
\ben
&&\partial_\rho \left(\sinh \bigg( \frac{\rho}{\sqrt{2k}} \bigg) \cosh \bigg( \frac{\rho}{\sqrt{2k}} \bigg) \partial_\rho \, a \right) + \frac{1}{2k} \, \frac{\cosh^3\big( \frac{\rho}{\sqrt{2k}} \big)}{\sinh \big( \frac{\rho}{\sqrt{2k}} \big)} \partial^2_\theta \, a \nonumber \\
&&\,- \sinh \bigg( \frac{\rho}{\sqrt{2k}} \bigg) \cosh \bigg( \frac{\rho}{\sqrt{2k}} \bigg) \partial^2_t \, a=0, \label{eom}
\een
It is convenient to put it in  a canonical Schr\"{o}dinger form. This is done by defining $u(r) \equiv \sinh^{1/2}(r) \, a(r)$, where $r=\sqrt{\frac{2}{k}} \,\rho$. Plugging the on-shell condition (\ref{onshell}) for a mode with some angular momentum $\ell$ and energy $E$ gives, \cite{Dijkgraaf:1991ba},
\be
\bigg(-\partial_r^2 +  V(r) \bigg) u(r) = s^2 \: u(r),
\ee
with
\be
V(r) = \cosh^{-2} (r/2)  \bigg[ \bigg( m^2-\frac{1}{16} \bigg)\coth^2 (r/2) +\frac{1}{16}  \bigg]. \label{schrodinger}
\ee
This equation can be solved exactly in terms of hypergeometric functions, whose asymptotic behavior gives
\be
 R_{pert}  = \nu^{2j+1} \frac{\Gamma(m-j)}{\Gamma(m-(-j-1))} \, \frac{\Gamma(-\bar{m}-j)}{\Gamma(-\bar{m}-(-j-1))} \, \frac{\Gamma(1+2j)}{ \Gamma(1+2(-j-1))}. \label{Rpert}
\ee
$\nu$ is a constant that is related to the value of the string coupling at the tip. It is convenient to choose $\nu=1/4$ so that, as in standard QM, at high energies the phase shift becomes trivial. With this choice, we can expand the GR phase shift at high energies to find
\be\label{cla}
\delta_{pert}= -\pi (1/2 +|\ell|)+\frac{2 \ell^2-1}{4 s} + O(s^{-2}).
\ee
The first term is the trivial term in $\mathbb{R}^2$ that is induced by the centrifugal potential. When considering scattering in $\mathbb{R}^2$, this term is usually omitted. Here we consider scattering on the cigar and we keep it as  a useful reference to know at which energies $\mathbb{R}^2$ is a good approximation.

The second term is the leading curvature correction to $\mathbb{R}^2$. Near the tip, the potential (\ref{schrodinger}) agrees with the centrifugal potential in $\mathbb{R}^2$. The difference between the two is responsible for this second term in the phase shift. Equation (\ref{cla}) implies that this difference becomes important at distances of the order of $\sqrt{k}$ from the tip, which indeed is the distance at which the $\mathbb{R}^2$ approximation to the cigar breaks down.

\subsection{The exact reflection coefficient}

The way the reflection coefficient is calculated in the exact coset CFT is the following. We consider the   vertex operators on the cigar that correspond to $j=-\frac12 +i s$,
which asymptotically take the form (see e.g. \cite{Aharony:2004xn})
\be
 \,V_{j;m,\bar{m}} \sim \left( e^{i p \rho} + R_{CFT}(j;m,\bar{m}) \, e^{-i p \rho}   \right).
\ee
An overall $m$ and $\bar{m}$ dependent pre-factor, that plays no role here, was omitted.
The two-point-function of such operators, normalized such that the coefficient of their asymptotic incoming piece is 1,
gives the reflection coefficient $R_{CFT}(j;m,\bar{m})$.
One can calculate it via the original bootstrap approach \cite{Teschner:1999ug}
and/or with the help of screening charges \cite{Giribet:2000fy,Giribet:2001ft} to get eq. (\ref{ftg}),
where the non-perturbative correction to the reflection coefficient takes the form
\be\label{strcor}
R_{non-pert}=
\frac{\Gamma(1+\frac{2j+1}{k})}{\Gamma(1-\frac{2j+1}{k})}~.
\ee
{}From a target-space perspective, the term (\ref{strcor}) is interesting since, unlike  $R_{pert}$, it contains information that goes beyond the GR background (\ref{cigar}).
In particular, (\ref{strcor})
is controlled by a scale $l_s/\sqrt{k}$, which is much shorter than the GR scale,
$\sqrt{k} \, l_s$, in the large $k$ limit. Our main goal here is to extract the target-space meaning of (\ref{strcor}).

Naively, one may suspect that this goal cannot reveal dramatic modifications to GR,
as it seems that $R_{non-pert}$ is induced by standard irrelevant terms, and so is negligible compared to $R_{pert}$. In particular, $R_{non-pert}$ takes a similar form
to the last ratio of $\Gamma$ functions in $R_{pert}$, (\ref{Rpert}),
only with a much shorter scale. This is related to the exact calculation of the reflection coefficient in Liouvile theory \cite{Zamolodchikov:1995aa}. Thus, it may be hard to imagine that it could have a distinct effect. It turns out, though, that in the cigar geometry, unlike in Liouvile theory, this reasoning is misleading \cite{I}.

As discussed above, at high energies, the other four $\Gamma$ functions (that do depend on $m$ and $\bar{m}$) cancel the leading behavior of the other two $\Gamma$ functions
that appear in $R_{pert}$, so that we end up with  eq. (\ref{cla}).
This is the reason that the phase shift goes to zero at high energies. Or, more generally, if we do not fix $\nu$ in (\ref{Rpert}),
that the density of states, defined by
\be
\rho(s)=\frac{1}{2\pi i} \frac{d \log R}{ds}~,
\ee
goes to a constant at high energies.
This is a general result for scattering in QM. Hence, we expect any perturbative $\alpha'$ and $g_s$ corrections to have this property. Such corrections will modify the sub-leading terms in (\ref{cla}), but not the leading term. The reason is that this term merely reflects the fact that there is a centrifugal barrier at the tip, which is induced by the use of radial coordinates. Since  perturbative $\alpha'$ and $g_s$ corrections
do not render the tip  special, they cannot modify the leading behavior  in  (\ref{cla}).

However, as argued above, the origin of $R_{non-pert}$ is a non-perturbative $\alpha'$ effect, due to the condensation of a tachyon field,
associated with a wound string.
It was argued in \cite{Kutasov:2005rr}--\cite{I} that this non-perturbative mode does mark the tip as a special point, even at large $k$. Hence, it is not clear that $R_{non-pert}$ should yield a constant energy density at high energies. Put differently, if $R_{non-pert}$ modifies the leading behavior of $R_{pert}$, then it is very likely that the tip of the cigar is a special point, even at parametrically small curvature.

In \cite{I},
it was pointed out that at high energies, $p\gg\sqrt{k}$,
the phase shift induced by $R_{non-pert}$ is
\be\label{ji}
\delta_{non-pert}= 2 \, \sqrt{\frac{2}{k}} \, p \log(p) \gg 1~.
\ee
This means not only that $R_{non-pert}$ becomes the dominant factor, but that it modifies in a rather dramatic way the behavior of the GR phase shift (and energy density) at high energies. Some aspects of this drastic modification were discussed in \cite{I}. Here we elaborate further on its consequences and origin.

%

\section{UV/IR mixing}

So far, the extra time direction, $t$, was a spectator in the analysis; it was needed for allowing the on-shell condition (\ref{onshell}), but besides that it did not play any role. In this section, we study the dependence of the reflection coefficient on $t$, focusing on large time scales.

The time dependence of the reflection coefficient is obtained in the standard way. In the previous section, we discussed  $R_{\ell}(p)$. Using the on-shell condition in string theory, (\ref{onshell}), we can write the reflection coefficient as a function of energy, $R_{\ell}(E)$. The Fourier transform
of $R_{\ell}(E)$ is denoted by $f_{\ell}(t)$.
Since the reflection coefficient is obtained from a two-point-function calculation,
we can view  $f_{\ell}(t)$ as a correlator at some time separation $t$,
\be\label{hy}
f_{\ell}(t)\equiv\langle {\cal O}_{\ell}(t) {\cal O}_{\ell}(0) \rangle.
\ee
A motivation to inspect (\ref{hy}) comes from \cite{Maldacena:2001kr}, in which similar expressions were studied in the context of the AdS/CFT correspondence.
There, the behavior at exponentially large times can be viewed as an order parameter for the topology of thermal AdS. Some puzzles concerning the thermal CFT versus thermal AdS were raised in \cite{Barbon:2003aq}. Resolving these issues is likely to shed light on the information puzzle.

Here, however,  the situation  is quite different than in \cite{Maldacena:2001kr}.
First, we consider the Euclidean setup with an extra time direction
(the Lorentzian case will be discussed elsewhere \cite{fu}).
Second, unlike in AdS/CFT, here we have a continuum at infinity that renders the Poincar\'e recurrence argument, discussed in \cite{Maldacena:2001kr,Barbon:2003aq},
irrelevant. Still,  we have an exact result and it is interesting to see what its Fourier transform gives. As we shall see, it is much more dramatic than Poincar\'e recurrence.

The important features of $f_{\ell}(t)$  do not depend on $\ell$.
The reason is that, as is clear from (\ref{onshell}), $\ell$ is relevant at low energies, while the effect we wish to describe is due to high energies. Hence, for simplicity, instead of working with (\ref{onshell}), we take $E=p$,
and denote the result of the Fourier transform by  $f(t)$.

Since we focus on the long-time behavior of $f(t)$,
one may suspect that the non-perturbative correction plays no role.
More precisely, the relevant energies for the non-perturbative correction are
of the order of $\sqrt{k}/l_s$.
Hence, they are expected to be negligible for $t\gg l_s/\sqrt{k}$.
Intriguingly, this turns out to be wrong. Below, we show that, in fact, the
correction due to the winding string condensate becomes more and more dominant as we increase $t$.
In other words, this is an example of UV/IR mixing: from energy point of view,
the non-perturbative correction affects only the UV,
but it modifies drastically the IR in time.


We shall begin by considering the Fourier transform of the reflection coefficient in GR,
$R_{pert}$. Then, as an educational warm-up exercise, we consider the Fourier transform of
the piece arising due to the winding string condensate, $R_{non-pert}$. Finally,
we address the Fourier transform of the exact reflection coefficient, $R_{CFT}$.


\subsection{The Fourier transform of $R_{pert}$ }


Using $\Gamma$-functions identities, the relevant expression  reads
\be
f_{pert} (t) =  \frac{1}{2\pi} \int_{-\infty} ^\infty dE \, e^{-i E \left(t -4\sqrt{\frac{k}{2}} \log(2)\right)} \, \frac{\Gamma \left(-2i\sqrt{\frac{k}{2}} E \right)}{\Gamma \left(2i\sqrt{\frac{k}{2}} E \right)}  \left( \frac{\Gamma \left(i\sqrt{\frac{k}{2}} E \right)}{\Gamma \left(-i\sqrt{\frac{k}{2}} E \right)} \right)^2~, \label{2ptFTcl}
\ee
where we used
the choice of normalization, $\nu=1/4$, and the simplifying $p=E$ framework.
This  integral can be performed exactly (see appendix).
Here, we just discuss some aspects of its asymptotic time behavior.

{}For positive $t$, we close the contour of integration
with an arc of infinite radius on the lower half of the complex plane,
and so the large $t$ behavior is controlled by the nearest pole in the lower-half plane, which gives
\begin{equation}
f_{pert} (t\to\infty) = -\frac12 \sqrt{\frac{2}{k}}\,  e^{-\frac12 \sqrt{\frac{2}{k}} \, t}.
\label{largetgr}
\end{equation}
{}For negative $t$, we close the contour with an arc of infinite radius on the upper half of the complex plane, and so the large $t$ behavior is controlled by the nearest pole in the upper-half plane, which gives
\begin{equation}
f_{pert} (t\to -\infty) = \frac14 \sqrt{\frac{2}{k}}\,  e^{ \sqrt{\frac{2}{k}} \, t}.
\end{equation}
To recapitulate, we see that in GR,
while there is an asymmetry between negative and positive separation,\footnote{The
reason for the asymmetry is that we calculated
$\int_{-\infty}^{\infty} dE \exp(-iEt)\, R(E)$, that is the same as
$Re\left(\int_{0}^{\infty} dE \exp(-i E t)\, R(E)\right)$,
which is not invariant under $t\to-t$;
an integral that is invariant under time reversal is e.g.
$Im\left(\int_{0}^{\infty} dE \exp(-i E t) R(E)\right)$.}
the result on both sides vanishes exponentially fast asymptotically in time,
as expected.

\subsection{The Fourier transform of $R_{non-pert}$ }

Next, we turn to the Fourier transform of $R_{non-pert}$.
As we show, this is a useful exercise, prior to calculating
the Fourier transform of the full reflection coefficient.
The relevant integral reads
\be
f_{non-pert} (t) =   -\frac{1}{2\pi} \int_{-\infty} ^\infty dE \, e^{-i E t } \, \frac{\Gamma \left(i\sqrt{\frac{2}{k}}E \right)}{\Gamma \left(-i\sqrt{\frac{2}{k}}E \right)}~. \label{2ptFTst}
\ee
This integral, too, can be solved exactly\footnote{It
is closely related to the kernel calculated in \cite{Natsuume:1994sp}.}
(see appendix);
here, we focus on educational properties concerning its asymptotic time behavior.

There are two important related differences between this integral and the GR case;
both are due to the fact that  at high energies
\be
R_{non-pert} \sim \exp\left( i \sqrt{8/k} \, E \log E \right) .
\ee
First, this means that if we wish to perform the integration with the help of the residue theorem then,  regardless of the sign of $t$, we are forced to close the contour on an arc of infinite radius on the upper half of the complex plane. Second, the asymptotic behavior of
$R_{non-pert}$ also means that at large and positive $t$ there is a
saddle point. We can write the integrand as a phase, and check when it is stationary.
{}For $t \gg \frac{1}{\sqrt{k}}$, the equations become simple, and we find that
there are two saddle points, at $E=\pm E_{sp}$, where
\begin{equation}
E_{sp}=\sqrt{\frac{k}{2}} \, e^{\frac12 \sqrt{\frac{k}{2}}\, t}. \label{spstonly}
\end{equation}
This saddle point is exponentially large in $t$, and so it gives contribution only from very high energy. Equation (\ref{spstonly}) is, in fact,
the first hint for UV/IR mixing -- the
{\it stringy correlation functions at large $t$ are dominated by high energy modes}.

The saddle-point approximation gives,
for  large and positive $t$,
\be
f_{non-pert} (t\to\infty) = -\sqrt{\frac{k}{2\pi}} \,  e^{\frac14 \sqrt{\frac{k}{2}} \, t} \cos \left(2  e^{\frac12 \sqrt{\frac{k}{2}} \, t}   +\frac{\pi}{4}\right). \label{2ptFTstresasymp}
\ee
This result is vastly different from the GR behavior.
First, the amplitude {\it grows} exponentially with $t$.
Second, it oscillates wildly.
As we shall see, the fact that it oscillates faster than the amplitude grows
has important consequences.

\subsection{The Fourier transform of $R_{CFT}$}

At last, we are ready to analyze the Fourier transform of the full reflection coefficient, $R_{CFT}$,
\be
f(t) = - \frac{1}{2\pi} \int_{-\infty} ^\infty dE \, e^{-i E \, \left( t-4 \sqrt{\frac{k}{2}} \log (2)\right)} \, \frac{\Gamma \left(-2i \sqrt{\frac{k}{2}} E \right)}{\Gamma \left(2i\sqrt{\frac{k}{2}} E \right)} \frac{\Gamma \left( i\sqrt{\frac{2}{k}} E \right)}{\Gamma \left( -i\sqrt{\frac{2}{k}} E \right)} \left( \frac{\Gamma \left(i\sqrt{\frac{k}{2}} E \right)}{\Gamma \left(-i\sqrt{\frac{k}{2}} E \right)}\right)^2. \label{2ptFT}
\ee
If we wanted to find the exact form of $f(t)$,
then by the same reasoning as in the previous subsection,
we had to close the contour with an arc at infinity on the upper-half plane.

\begin{figure}
\centerline{\includegraphics[scale=0.47]{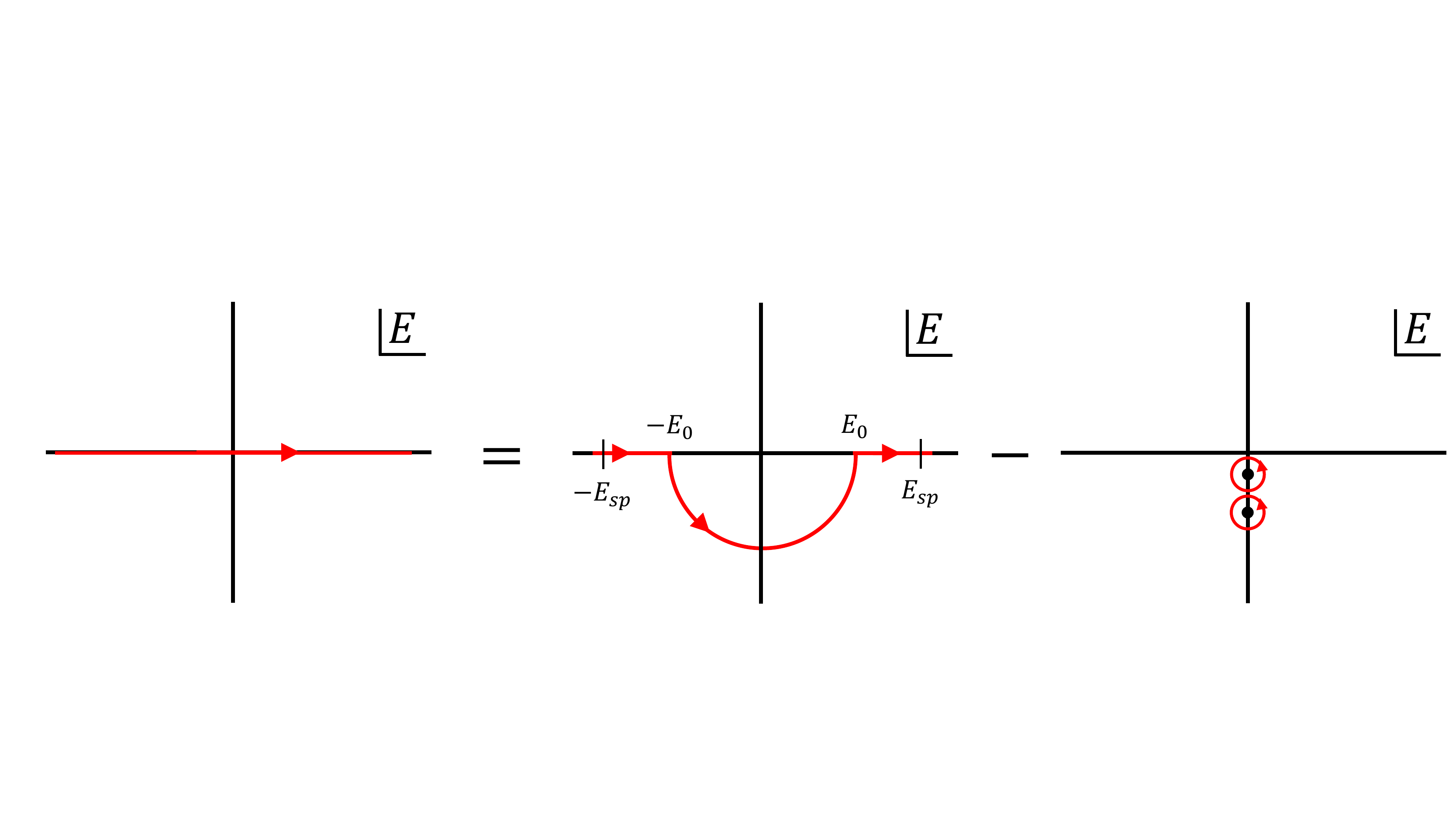}}
\caption{We deform the integral on the real axis and subtract the residue contributions of the poles that crossed the contour due to the deformation. The integral over the arc in the first term on the right hand side is very small, and so the entire term is approximately equal to the saddle-point contribution alone.}
\label{contourdeform}
\end{figure}

Our goal here is more modest; we seek to find $f(t)$ only at large and positive $t$.~\footnote{At large and negative $t$, the non-perturbative correction leads to a negligible correction to the perturbative result.}
With that goal in mind, a convenient contour is the one described in figure \ref{contourdeform}.
On the left of figure \ref{contourdeform}, we have the contour of the integral we are after. By the residue theorem, it is equal to the integral over the contour that appears on the right of the equality minus the residues, as indicated in the figure. For large and positive $t$, the contribution of the arc in the first term on the right of
figure \ref{contourdeform} is negligible, and the straight lines are dominated, as in the previous subsection, by the saddle point, which gives
\be
f(t\to\infty) = - \sqrt{\frac{k}{2 \pi}}  \, e^{\frac{1}{4} \sqrt{\frac{k}{2}}\,t} \cos \left( 2  e^{\frac{1}{2} \sqrt{\frac{k}{2}}\,t} + \frac{\pi}{4} \right). \label{2ptFTsaddle}
\ee
This part is identical to the stringy-only result for large $t$, (\ref{2ptFTstresasymp}). 

At large $t$, the dominant contribution from the poles comes from the first pole,
whose contribution is, up to tiny corrections, well approximated by (\ref{largetgr}).
Therefore, at large $t$, we find that~
\be
f(t\to\infty)= - \sqrt{\frac{k}{2 \pi}}  \, e^{\frac{1}{4} \sqrt{\frac{k}{2}}\,t} \cos \left( 2  e^{\frac{1}{2} \sqrt{\frac{k}{2}}\,t} \right)-\frac12 \sqrt{\frac{2}{k}}  \, e^{-\frac12 \sqrt{\frac{2}{k}} \, t}. \label{2ptFTsaddlecor}
\ee
Here we ignored the constant phase in the first term,
which is unimportant at large $t$ and/or large $k$.
The reason we care about the second term,
which is due to the large $t$ behavior of the GR piece,
although it appears negligible compared to the first one,
is that it does not oscillate with time. As we shall show below,
this is crucial when we coarse grain this expression.\footnote{Using the
exact on-shell condition -- (\ref{onshell}) with generic,
finite angular momentum $\ell$ (instead of $E=p$),
would only modify the large $t$ behavior of the GR piece (the second term)
and not the universality of the first term (the one due to the winding string correction).
}

Comparing $f(t)$ with $f_{pert}(t)$, it is evident that the correction due to the winding string condensate induced UV/IR mixing. The large $t$ behavior of $f(t)$ is drastically different than that of $f_{pert}(t)$ (see figure 3).
Technically, the reason for that is the fact that the stringy correction affected the UV (in the energy sense) in such a dramatic way, that it generated a saddle point that does not exist in GR. In the next subsection, we discuss the physical origin of this saddle point and its generality.

\begin{figure}
\centerline{\includegraphics[scale=0.47]{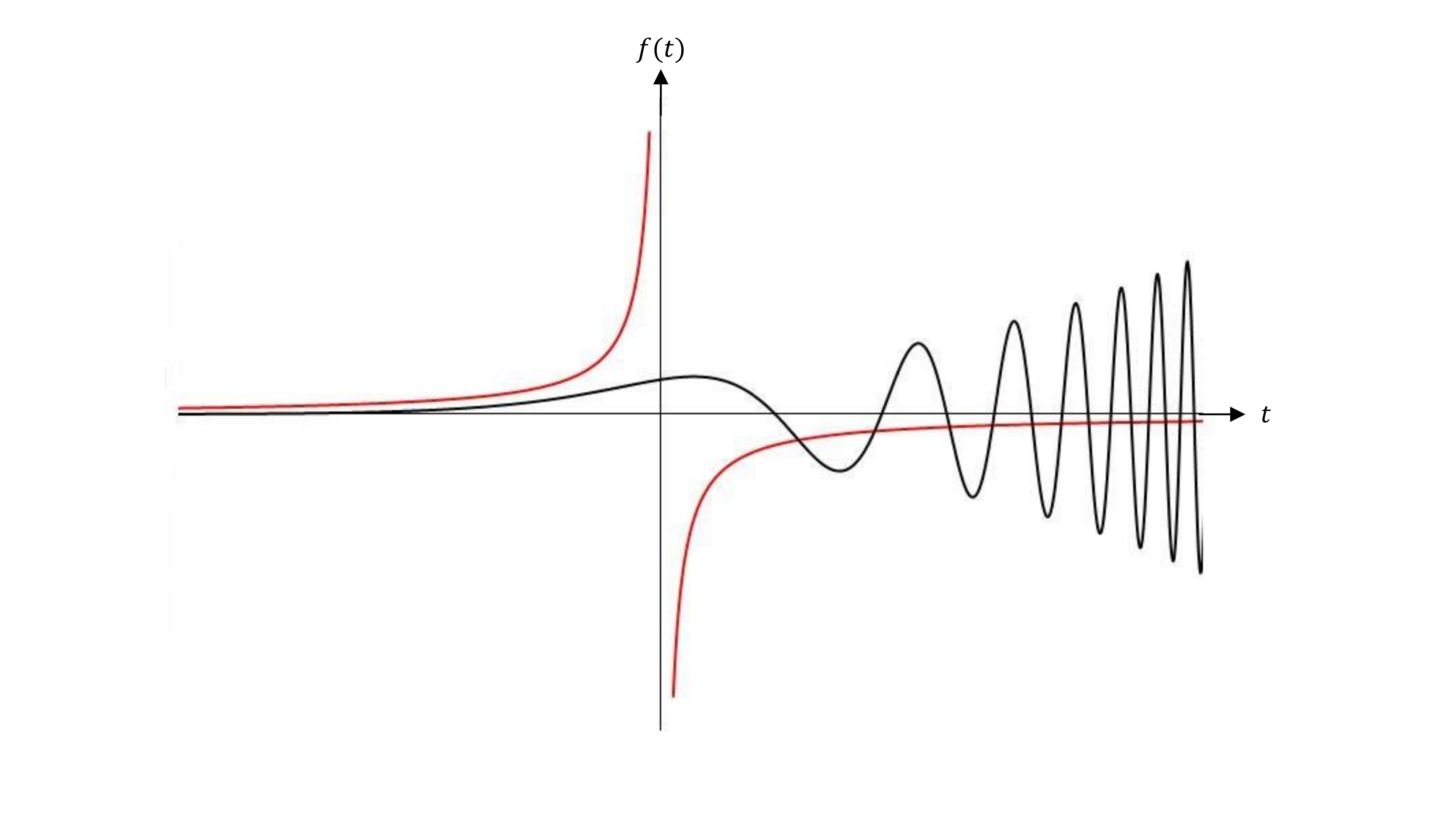}}
\caption{The perturbative (red) and non-perturbative (black) contributions are plotted. At large and negative $t$, the non-perturbative contribution is negligible. However, at large and positive $t$, it is dominant.}
\label{comparison}
\end{figure}

\subsection{ Origin of the UV/IR mixing and generality}

{}From a mathematical stand point, the origin of the UV/IR mixing is the fact that
the non-perturbative correction induced a saddle point in the integral that defines $f(t)$, which is simply absent in the perturbative,
GR expression. In this subsection, we discuss the physical origin of this saddle point and its generality.

The physical process that we considered involves sending a wave from infinity that spends some time at the cap of the cigar -- the region where the curvature is $\sim 1/\sqrt{k}$ -- before scattering back to infinity. The two-point-function is sensitive to the amount of time the wave spends at the cap. In GR, this time is finite for any energy $E$, including  the limit $E\to\infty$. As a result, at positive and large $t$, we merely see the tail of the wave function that drops exponentially.

The winding string correction changes this in a dramatic way.
The full phase shift implies that as we increase the energy,
the wave is spending more and more time at the cap before scattering back to infinity.
The exact CFT phase shift implies that for any time separation, no matter how large,
there is an energy $E(t)$ such that the wave is spending just the right amount of
time at the cap to scatter to infinity at $t$. Since this happens at very high energies, the wave propagation is well approximated by a massless particle trajectory (see figure 6). The fact that the wave propagation is well approximated by this particle trajectory
is the physical origin of the saddle point in the integral.

In the next section, we elaborate on this particle trajectory, and address natural questions such as:
where in the cap does the wave spend the extra time? And, what happens at lower energies? Below, we address a different question: how general is the UV/IR mixing that we discussed?

Although we presented a concrete example -- the $SL(2)_k/U(1)$ SCFT,
we believe that our results are more general.
What goes into our derivation are the following facts:  (a) At high energies the GR/perturbative reflection coefficient goes to $1$. (b) It is dressed by a non-perturbtive factor,
$R_{non-pert}(E)$, with an $E\log(E)$-type behavior. As discussed in section 2, we expect (a) to be generic and, in particular, to hold in Schwarzschild black holes in four dimensions. (b) is more subtle. The fact that the full  reflection coefficient can be written as a product of the form (\ref{ftg}) is a non-trivial property of
the coset CFT, $SL(2, \mathbb{R})_k/U(1)$, and it is unlikely to be generic. However, the origin for the large phase shift in
$SL(2, \mathbb{R})_k/U(1)$ is the condensation of the wound string at the tip. Since this condensation \cite{Kutasov:2005rr} and its features
\cite{Giveon:2012kp,Giveon:2013ica,Giveon:2014hfa} are believed to be general
-- for any cigar-like CFT background,
it is natural to suspect that the large phase shift and the UV/IR mixing are generic
as well, in such cases. Needless to say that it would be nice to make this concrete.

\section{Effective description of the stringy correction}

The discussion so far used information that is natural in string theory
-- the S-matrix. This S-matrix information  illustrates  perplexing features, but it does not reveal its meaning. In particular, we showed  that even for large $k$ and at large $t$ separation, the background (\ref{cigar}) is missing important physics, since it naturally leads to the second term in (\ref{2ptFTsaddlecor}), but not the dominating first term of (\ref{2ptFTsaddlecor}).

In \cite{I}, we showed that, as it is often the case in string theory \cite{Gross:1987kza}, at large $E$ the string theory S-matrix is dominated by a saddle point. The target-space shape of this saddle point is an additional useful information. In \cite{I}, it was shown  that the shape of the saddle point goes beyond the tip of the cigar into a region that simply does not exists in (\ref{cigar}).

In this section, we study the target-space meaning of the CFT two-point-functions using an effective description. We do so both at low ($p\ll \sqrt{k}$) and high ($p\gg \sqrt{k}$) energies. Note that a ``low" energy can still be much larger than the curvature scale, $1/\sqrt{k}$, and even the string scale, $1$.
The effective description we use treats the string as a point particle. The justification to use this approximation comes from the observation of \cite{I} that it is the zero-mode integration in the full stringy problem that is responsible for the large phase-shift.

\subsection{Low energy}

\begin{figure}
\centerline{\includegraphics[scale=0.34]{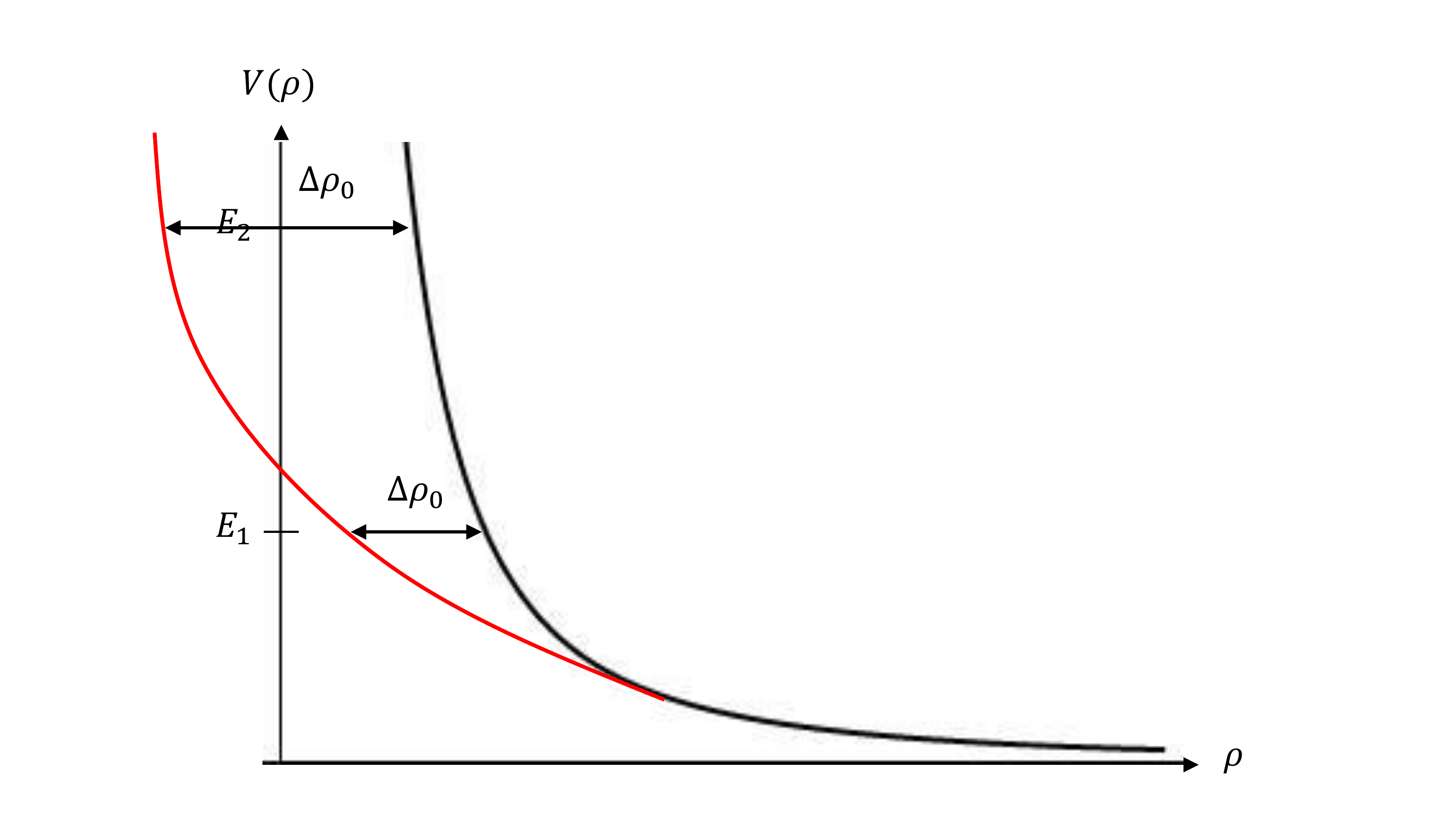}}
\caption{A plot of the effective stringy correction to the centrifugal potential. To leading order, the addition of phase shift $\Delta \delta_\ell$ pushes the turning point $\rho_0$ so that $\Delta \rho_0 = \Delta \delta_\ell /p$. At low energies, the stringy phase shift is small and we can find an effective potential close to the GR one. To account for the phase shift at high energies, the turning point is at negative $\rho$ -- a region that does not exist classically.}
\label{wkbturningpoint}
\end{figure}

At low energy, we can expand the phase
$R_{non-pert}(s/k)\equiv e^{-i\delta_{non-pert}}$, (\ref{strcor}), in powers of $s/k$,
\be
\delta_{non-pert}=c_1 s/k +c_3(s/k)^3+...~.
\ee
The linear term can be absorbed into the definition of $\nu$,
so the first interesting correction is the cubic term, for which
$c_3=\frac23 \zeta(3)$ and, therefore,
\be\label{e}
\delta_{non-pert} \sim \frac{p^3}{k^{3/2}}~.
\ee
Note that the sign in (\ref{e}) is positive. The first question we wish to address is:
when does $\delta_{non-pert}$ becomes important compared to $\delta_{pert}$?
We consider energies that are low compared to $\sqrt{k}$, but are much larger than   $1/\sqrt{k}$, so that the expansion (\ref{cla}) is valid. We see that at low energies, the trivial part of (\ref{cla}) always dominates $\delta_{non-pert}$, and that for
\be
p^2\gg \sqrt{k} \, |\ell|, \label{condition}
\ee
$\delta_{non-pert}$ is larger than the curvature contribution to $\delta_{pert}$. Thus, in the large $k$ limit (with a fixed $\ell$), there is a  range
\be\label{range}
\sqrt{k} \gg p \gg k^{1/4},
\ee
in which the stringy correction, and not the cap curvature, is the leading deformation of $\mathbb{R}^2$. In this range, the density of states becomes an increasing function of $p$. However, the $\mathbb{R}^2$ approximation still does not break, since the increase in the density of states can be accounted for by a modification of the centrifugal potential of $\mathbb{R}^2$ (see figure \ref{wkbturningpoint}).

To make this more precise, we can work in the ``Schr\"{o}dinger frame," in which the kinetic term is canonical, and ask what  modification to the centrifugal potential is required in order to get $\delta_{non-pert}$.
Since in the range (\ref{range}) $\delta_{non-pert} \ll 1$, we expect this modification, which we denote by $\Delta V$, to be much smaller than the centrifugal potential. Writing $V=V_{cen}+\Delta V$,
we can approximate $\Delta V$ using the leading term in the WKB approximation,
\be
\Delta V \sim \frac{\delta\phi}{p} \partial_\rho V_{cen}.
\ee
Since $V_{cen}=\ell^2/\rho^2$, we have $p=\ell/ \rho$, and so using (\ref{e}) we get
\be\label{cor}
\Delta V \sim - \frac{\ell^4}{k^{3/2} \, \rho^5}.
\ee
The minus sign implies that $V < V_{cen}$,
as it should (see figure \ref{wkbturningpoint}).
As expected, at low energies this is a small effect.
Indeed, $\Delta V$ is of the order of $V_{cen}$ at $\rho\sim k^{-1/2}$,
which in momentum space means the upper bound in (\ref{range}).

Since it blows up at the tip, (\ref{cor}) appears to mark the tip as a special point in the cigar geometry.
One may wonder if we missed a less radical explanation to $\delta_{non-pert}$ that does not mark the tip as a special point, because of the fact that we worked in the ``Schr\"{o}dinger frame." In particular, is it possible that $\Delta V$ is merely one term out of several irrelevant terms that sum up into a standard higher order correction to GR, such as an $R^2$ term? In such a case, it would not indicate that the tip is special.

Put differently, the centrifugal term is obtained from $\Box\Phi$, which clearly does not mark the tip as a special point. Can we similarly get (\ref{cor}) from some other, less relevant, terms?
Since (\ref{cor}) includes $\ell^4$, we can try $\Box^2 \Phi$. This, however, gives $\ell^4/\rho^4$ and not $\ell^4/\rho^5$. In order to obtain an additional factor of $1/\rho$, we can try $\partial_{\rho} \, \Box^2 \Phi$, but this is not a scalar. To turn it into one, we can try
$\partial^{\mu} {\cal O} \, \partial_{\mu} \Box^2 \Phi$,
with ${\cal O}$ a scalar like the curvature or dilaton. However, in the cigar background it appears that ${\cal O}$ is always a series of even powers of $\rho$, and so
it cannot give rise to (\ref{cor}).

This reasoning seems to imply not only that the stringy correction is non-perturbative in $\alpha'$ (as perturbative $\alpha'$ corrections do not mark the tip special), but also that it cannot be written as an irrelevant term in the standard Wilsonian approach. In fact, non-Wilsonian terms, such as the ones discussed in \cite{Itzhaki:2004dv}, appear to be needed here. It is tempting, therefore, to think of the wound string condensate
as the horizon order parameter anticipated in \cite{Itzhaki:2004dv}.

\subsection{High energy}

What happens when we probe the cigar  with energies much larger than $\sqrt{k}$?
Then we have (\ref{ji}),
which means that $\delta_{non-pert}$ becomes so large that, for fixed $\ell$, it dominates even the trivial term in $\delta_{pert}$. In other words, the modification to the centrifugal potential is so large that (\ref{ji}) cannot be accounted for in $\mathbb{R}^2$, as the required phase shift takes $\rho$ to be smaller than $0$ (see figure \ref{wkbturningpoint}). Hence, as the full stringy analysis implies \cite{I}, we have to consider extension of $\mathbb{R}^2$ beyond the origin.

The phase shift (\ref{ji}) can be reproduced by an effective exponential potential,
\be\label{lio}
V=c \, e^{-\sqrt{2k} \rho},~~ \mbox{with}~~c=2k.
\ee
Any positive value of $c$ is consistent with  (\ref{ji}). What determines $c$ is  $\delta_{non-pert}$ and the reference ``trivial" part of $\delta_{pert}$, meaning, the demand that sub-leading terms reproduce the phase $-\pi(1/2 + |\ell|)$.

This effective potential is the same as the radial potential in Sine-Liouville theory
that is FZZ-dual to the cigar theory \cite{fzz,kkk}. We see that the effective description leads  to the same conclusion as obtained by the full string theory analysis \cite{I},
namely, that at high energies the coset $SL(2, \mathbb{R})_k/U(1)$ is better described by the Sine-Liouville theory than by the cigar geometry.

\begin{figure}
\centerline{\includegraphics[scale=0.44]{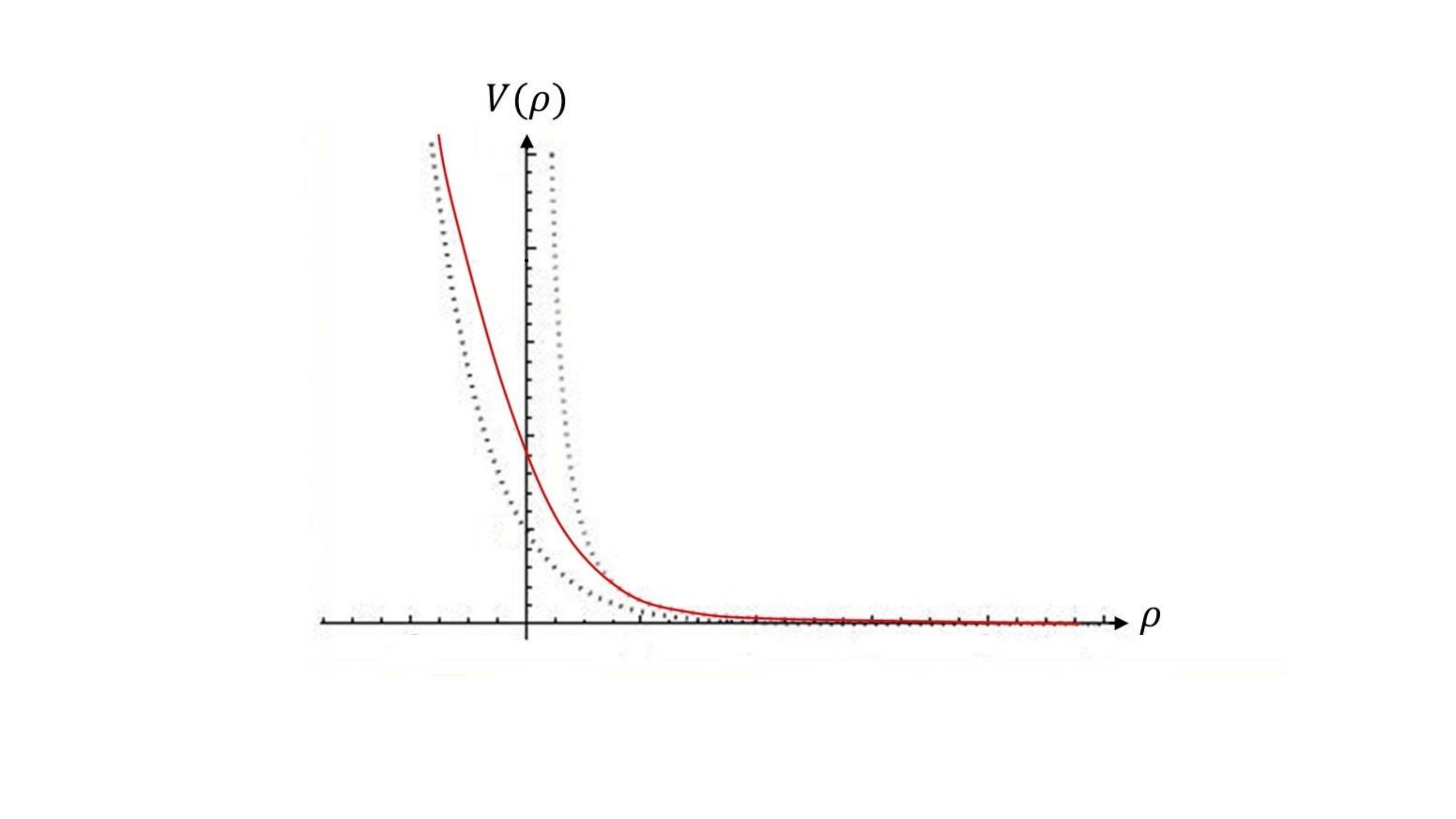}}
\caption{The full potential at high energies. The red curve is a natural interpolation between two asymptotic curves: the dotted line on the right is asymptotic at $\rho \to \infty$ and behaves like the classical centrifugal potential, $1/\rho^2$, while the dotted line on the left is asymptotic at $\rho \to -\infty$ and behaves like the Sine-Liouville potential, $e^{-\sqrt{2k} \,\rho}$.}
\label{fullpotential}
\end{figure}

The fact that $c\sim k$ goes well with the fact that the transition between the low and high energies takes place at $p\sim\sqrt{k}$. As discussed above, at that momentum the correction to the centrifugal barrier is of order $1$ (in units of the centrifugal barrier). The value of the centrifugal barrier at such energies is of order $k$, as is the value of (\ref{lio}) around $\rho=0$. It is natural to suspect, therefore, that the full potential is a monotonic decreasing function of $\rho$ that interpolates between the classical potential to the right and the Liouville potential to the left (see figure \ref{fullpotential}). Numerical simulations support this claim.

A closely related quantity was calculated in \cite{Giveon:2013ica}. There, with the help of \cite{Giveon:2001up} and \cite{Bershadsky:1991in}, the ratio between the canonically normalized semi-classical wound tachyon mode and the canonically normalized semi-classical dilaton mode was found to scale like $k$ in the large $k$ limit. This goes well with $c\sim k$, since the wound tachyon is the origin of the large phase shift.

We have just concluded that in string theory the tip of the cigar is effectively
glued  to a Sine-Liouville space.
It would be nice to know how big this gluing region is.
To answer this question, we evaluate the cross section, $\sigma(p)$, associated with $\delta_{non-pert}$. The calculation of quantities relevant to scattering in 2+1 dimensions is done in a manner (reviewed e.g. in \cite{Adhikari:1986}) very similar to 3+1 dimensions. We write the boundary condition for the wave function as
\be
\Psi (\rho, \theta)=e^{i (p_x x + p_y y)} + \sqrt{i/p} \, f(\theta) \, \frac{e^{i p \rho},}{\sqrt{\rho}}~.
\ee
Similar manipulations as in 3+1 dimensions give
\be\label{fth}
f(\theta)=\sqrt{2/\pi}\left( e^{i \delta_0} \sin(\delta_0) +2\sum_{\ell=1}^{\infty}\cos(\ell \,\theta) e^{i \delta_\ell} \sin(\delta_\ell) \right),
\ee
and the total cross section is
\be\label{cross}
\sigma(p)=\frac{4}{p}\left(  \sin^2 (\delta_0) +2\sum_{\ell=1}^{\infty} \sin^2 (\delta_\ell) \right).
\ee
At high energies, the phase shift $\delta_\ell\gg 1$ and does not depend on $\ell$,
and so we get
\be
\sigma(p)\sim \frac{\ell_{max}}{p}~,
\ee
where $\ell_{max}$ is the bound on the validity of the $\mathbb{R}^2$ approximation that we are using. A reasonable way to estimate $\ell_{max}$ is to see at what $\ell$ the curvature correction to the trivial phase shift of $\mathbb{R}^2$ is of the order of $\delta_{non-pert}$. This gives $\ell_{max}\sim p \sqrt{\log(p)}$, which in turn implies that
\be
\sigma(p)\sim \sqrt{\log(p)}~.
\ee
Hence, up to logarithmic effects, the size of the defect at the tip is order $1$ in stringy units, in agreement with \cite{I}. The logarithmic term can probably be attributed to the fact that we can penetrate deeper and deeper into the Sine-Liouville defect as we increase the energy.

With these results in hand, we can make clearer the statement that the physical source of the saddle point is a high-energy particle trajectory. For a particle with $\ell=0$, we neglect the $\theta$ direction in the target space and are left with an effective 1+1 dimensional space that is equivalent to Minkowski space-time and is presented in figure 6.
At  high enough energies, a massless particle can propagate for some time $\Delta t$ into the ``Liouville space'' that does not exist in GR.
The time a mode with energy $E$ spends on the Liouville side is
$\Delta t \sim \frac{1}{\sqrt{k}} \log(E)$,
meaning that more energetic particles will scatter to radial infinity at later times, giving an intuitive understanding of the origin of the UV/IR mixing shown previously.

\begin{figure}
\centerline{\includegraphics[scale=0.35]{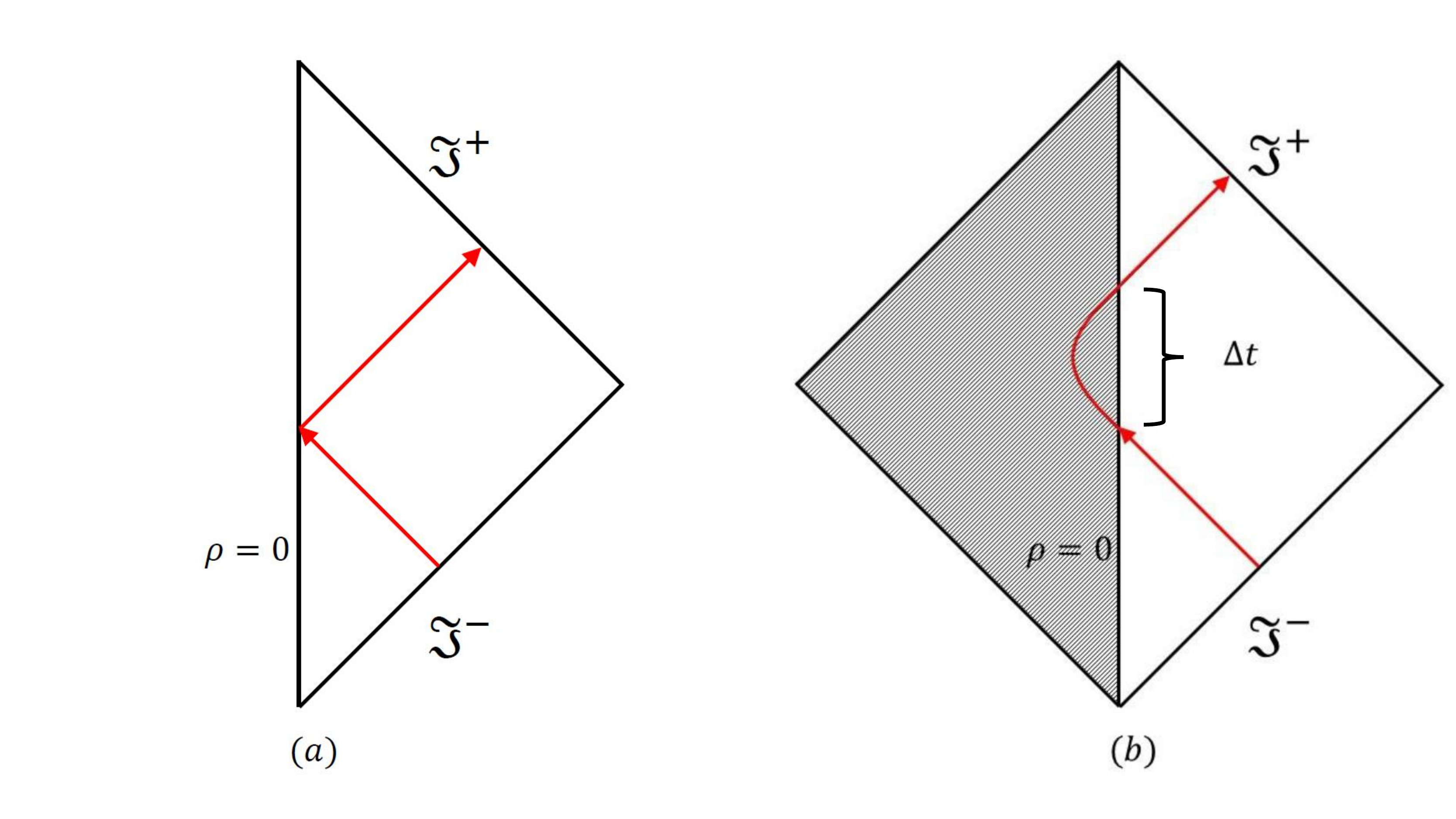}}
\caption{Diagram (a) shows the trajectory of a massless particle in the classical theory, or of a low-energy mode in the theory that includes non-perturbative corrections. Diagram (b) shows the trajectory of a massless particle with high energy in the full theory. The particle spends a time $\Delta t$ at the origin and thus connects points on $\mathscr{I^-}$ and $\mathscr{I^+}$ with a large time separation.}
\label{penrose}
\end{figure}

\section{Coarse graining}

In the previous sections,
we saw that at large $t$ separation the non-perturbative correction modifies the two-point-function of GR in a drastic way,
even for large $k$. In ``reality," there is always some uncertainty in $t$, which we denote $\delta t$. In this section, we show that unless $\delta t$ is exponentially small, at large $t$, it washes away completely the non-perturbative effects. Concretely, up to tiny corrections,
\be\label{app}
f_{\delta t}^{coarse-grain}(t) = f_{pert}(t),
\ee
where we define the coarse-grained Fourier transform in the following way:
\be
f_{\delta t}^{coarse-grain}(t) \equiv \frac{1}{\delta t} \int_t ^{t+\delta t} f(t') dt'.
\ee
The reason this averaging has such a drastic effect is simple.
Indeed, the non-perturbative contribution to $f(t)$ grows exponentially fast with $t$, but, unlike $f_{pert}(t)$, it also oscillates with time. As can be seen from (\ref{2ptFTsaddlecor}), it  oscillates much faster than the amplitude grows. For
\be\label{con}
\delta t \gg e^{-\frac{1}{2} \sqrt{\frac{k}{2}}\,t},
\ee
the averaging over the oscillations leads to a signal that goes like
\be\label{oa}
e^{\frac{1}{4} \sqrt{\frac{k}{2}}\,t}e^{-\frac{1}{2} \sqrt{\frac{k}{2}}\,t} = e^{-\frac{1}{4} \sqrt{\frac{k}{2}}\,t};
\ee
the first term on the left hand side is due to the exponential growth of the amplitude and the second term is due to the averaging over the oscillations.
We see that the left over, after coarse graining the stringy effect,
vanishes much faster than $f_{pert}(t)$, which means (\ref{app}).

This is another aspect of UV/IR mixing: $f_{\delta t}^{coarse-grain}(t)$ is extremely sensitive to the ratio
\be
r \equiv\frac{\sqrt{k}\,t}{|\log(\delta t)|}~.
\ee
{}For $r\ll 1$, we get $f_{pert}(t)$, while for $r\gg 1$, we get $f(t)$.
However, the fact that $\delta t\ll 1$ is  a UV scale and $t\gg 1$ an IR scale
does not fix $r$.

\section{Discussion}

We believe that the UV/IR mixing investigated here,
could improve our understanding of both
the black-hole information paradox 
as well as features of Little String Theory
(see e.g. \cite{Aharony:2004xn} and references therein).
We discuss each, in turn.

It is widely believed that, quantum mechanically, the black hole does not lose information and that the information is encoded in ``quantum hair." When we coarse-grain over the quantum states of the black hole, we conclude that the information is lost. Something similar is happening here. The non-perturbative effects in $l_s$ dominate the perturbative ones, as long as we are able to specify $t$ with an exponentially good accuracy. The longer we wait the better the accuracy has to be so that we see the ``stringy hair" associated with $f(t)$. This is very likely to remain a feature of Lorentzian black holes \cite{fu}.

Finally,
LST may be defined as the holographic dual of string theory on asymptotically linear-dilaton backgrounds \cite{Aharony:1998ub,Giveon:1999zm}.
In weakly coupled LST, we should introduce a wall in the strong coupling regime.
Whenever such a wall is described by a cigar-like sigma-model,
it falls into the generic class that we have in mind in this work.
In particular, in the DSLST examples in \cite{Giveon:1999px,Giveon:1999tq,Aharony:2004xn}, where (orbifolds of) the $SL(2)/U(1)$ SCFT times a {\it real} time appear explicitly,
the manipulations above apply automatically.
{}From this perspective, the UV/IR mixing we encountered
is a manifestation of the non-locality of LST.

\vspace{10mm}

\nl{\bf Acknowledgments}\\
We thank Eliezer Rabinovici and especially David Kutasov for many useful discussions.
The work of AG is supported in part by the BSF --
American-Israel Bi-National Science Foundation. The work of AG and NI is supported in
part by the I-CORE Program of the Planning and Budgeting Committee and the Israel
Science Foundation (Center No. 1937/12), and by a center of excellence supported by
the Israel Science Foundation (grant number 1989/14).

\begin{appendix}
\section{Fourier transform of $R$ }

\subsection{The Fourier transform of $R_{pert}$}
We solve the integral
\be
f_{pert} (t) =  \frac{1}{2\pi} \int_{-\infty} ^\infty dE \, e^{-i E \,t} e^{ 4i E\sqrt{\frac{k}{2}} \log(2)} \, \frac{\Gamma \left(-2i\sqrt{\frac{k}{2}} E \right)}{\Gamma \left(2i\sqrt{\frac{k}{2}} E \right)}  \left( \frac{\Gamma \left(i\sqrt{\frac{k}{2}} E \right)}{\Gamma \left(-i\sqrt{\frac{k}{2}} E \right)} \right)^2, \label{appsim2ptcl}
\ee
where the second exponential is compatible with the normalization factor, $\nu=1/4$,
and we simplified the expression with some $\Gamma$-function identities.

We perform the integral by closing a contour of the integration path and summing the pole residues. The contour we choose depends on the sign of $t$ -- for $t>0$ it is easy to show that the integrand goes to zero on an arc of infinite radius in the lower-half plane, while for $t<0$ it vanishes on an arc in the upper half of the complex plane.

We begin with $t<0$, and close the contour on an infinite semi-circle on the upper half of the plane.
In the upper-half plane, there are poles at $E_n=i \sqrt{\frac{2}{k}} \, n$, with $n$ a positive integer. The residue of the $n$-th pole is
\begin{equation}
\text{Res} |_{E_n=i \sqrt{\frac{2}{k}} \, n} = \frac{2^{-4n}}{i \pi } \, \sqrt{\frac{2}{k}} \, \frac{\Gamma(2n) \, \Gamma(2n+1) }{\Gamma^2 (n) \, \Gamma^2(n+1)} \, e^{\sqrt{\frac{2}{k}} \,t},
\end{equation}
and summing over $n$ gives
\be
f_{pert} (t<0) =\frac{1}{4} \,\sqrt{\frac{2}{k}} \,  e^{\sqrt{\frac{2}{k}}\, t } \, {}_2 F_1 \left( \frac32; \frac32; 2; e^{\sqrt{\frac{2}{k}}\, t }\right) . \label{appsim2ptpert-}
\ee
To find the asymptotic behavior, we can either take the residue of the first pole alone, or expand (\ref{appsim2ptpert-}) for large $t$. Both ways give
\begin{equation}
f_{pert} (t\to -\infty) = \frac14 \sqrt{\frac{2}{k}}\,  e^{ \sqrt{\frac{2}{k}} \, t}.
\end{equation}

For $t>0$, we close the contour in the lower half of the complex plane and get contributions from poles at $E_n=- \frac{i}{2} \sqrt{\frac{2}{k}} \, n$, with $n$ a positive and odd integer. In this case, the residues are
\begin{equation}
\text{Res} |_{E_n=- \frac{i}{2} \sqrt{\frac{2}{k}} \, n} = \frac{2^{-2+2n}}{i \pi } \, \sqrt{\frac{2}{k}} \, \frac{\Gamma^2(n/2) }{\Gamma (n) \, \Gamma(n+1) \, \Gamma^2 (-n/2)} \, e^{-\sqrt{\frac{1}{2k}} \,t},
\end{equation}
and summing them gives
\begin{equation}
f_{pert}(t>0) = -\frac{1}{2} \, \sqrt{\frac{2}{k}} \,  \frac{e^{\frac12 \sqrt{\frac{2}{k}}\,t} \, {}_2 F_1 \left( -\frac12; \frac12; 1; e^{-\sqrt{\frac{2}{k} \,t}} \right)}{e^{\sqrt{\frac{2}{k}}\,t}-1}. \label{appsim2ptpert+}
\end{equation}
Again, we can estimate the behavior at long times either from the first pole residue or directly from (\ref{appsim2ptpert+}), to get
\be
f_{pert} (t\to\infty) = -\frac12 \sqrt{\frac{2}{k}}  \, e^{-\frac12 \sqrt{\frac{2}{k}} \, t}. \label{appsim2ptpertasymp}
\ee

\subsection{The Fourier transform of $R_{non-pert}$}

The non-perturbative part of the transformed reflection coefficient is given by
\be
f_{non-pert} (t) =  -\frac{1}{2\pi} \int_{-\infty} ^\infty dE \, e^{-i E \,t}  \frac{\Gamma \left(i\sqrt{\frac{2}{k}} E \right)}{\Gamma \left(-i\sqrt{\frac{2}{k}} E \right)} , \label{appsim2ptst}
\ee
where again we simplified the expression with $\Gamma$-function identities. It is straightforward to check that at high energies, the $\Gamma$ functions dominate the integrand and so it vanishes only on a semi-circle of infinite radius on the
upper-half plane, regardless of the sign of $t$. The residues of the poles are of the form
\begin{equation}
\text{Res} |_{E_n= i \sqrt{\frac{k}{2}} \, n} = \frac{1}{2 \pi i } \, \sqrt{\frac{k}{2}} \, \frac{(-1)^{n-1}}{n!  \, (n-1)!} \, e^{\sqrt{\frac{k}{2}} \, n\, t},
\end{equation}
where $n$ is a positive integer. Summing over the residues gives
\begin{equation}
f_{non-pert} (t) = \sqrt{\frac{k}{2}} \, e^{\frac12 \sqrt{\frac{k}{2}} t} \, J_1 \left(2 \, e^{\frac12 \sqrt{\frac{k}{2}} t} \right),
\end{equation}
where $J_1 (z)$ is the Bessel function of order 1. Expanding for large separation through the use of the Bessel-function asymptotic form gives
\begin{equation}
f_{non-pert} (t\to\infty) = -\sqrt{\frac{k}{2\pi}} \,  e^{\frac14 \sqrt{\frac{k}{2}} \, t} \cos \left(2  e^{\frac12 \sqrt{\frac{k}{2}} \, t}   +\frac{\pi}{4}\right). \label{appsim2ptstasym}
\end{equation}
This is, of course, exactly the same approximation that was found in section 3,
via the saddle-point approximation.

\end{appendix}


\begin{thebibliography}{24}

\bibitem{Bekenstein:1973ur}
  J.~D.~Bekenstein,
  ``Black holes and entropy,''
  Phys.\ Rev.\ D {\bf 7}, 2333 (1973).

\bibitem{Hawking:1974sw}
  S.~W.~Hawking,
  ``Particle Creation by Black Holes,''
  Commun.\ Math.\ Phys.\  {\bf 43}, 199 (1975)
  [Commun.\ Math.\ Phys.\  {\bf 46}, 206 (1976)].

\bibitem{Callan:1988hs}
  C.~G.~Callan, Jr., R.~C.~Myers and M.~J.~Perry,
  ``Black Holes in String Theory,''
  Nucl.\ Phys.\ B {\bf 311}, 673 (1989).


\bibitem{Kutasov:2005rr}
  D.~Kutasov,
  ``Accelerating branes and the string/black hole transition,''
  hep-th/0509170.

\bibitem{Giveon:2012kp}
  A.~Giveon and N.~Itzhaki,
  ``String Theory Versus Black Hole Complementarity,''
  JHEP {\bf 1212}, 094 (2012)
  [arXiv:1208.3930 [hep-th]].


\bibitem{Giveon:2013ica}
  A.~Giveon and N.~Itzhaki,
  ``String theory at the tip of the cigar,''
  JHEP {\bf 1309}, 079 (2013)
  [arXiv:1305.4799 [hep-th]].


\bibitem{Giveon:2013hsa}
  A.~Giveon, N.~Itzhaki and J.~Troost,
  ``The Black Hole Interior and a Curious Sum Rule,''
  JHEP {\bf 1403}, 063 (2014)
  [arXiv:1311.5189 [hep-th]].


\bibitem{Giveon:2014hfa}
  A.~Giveon, N.~Itzhaki and J.~Troost,
  ``Lessons on Black Holes from the Elliptic Genus,''
  JHEP {\bf 1404}, 160 (2014)
  [arXiv:1401.3104 [hep-th]].


\bibitem{I}
  A.~Giveon, N.~Itzhaki and D.~Kutasov,
  ``Stringy Horizons,''
  arXiv:1502.03633 [hep-th].

\bibitem{Mertens:2013pza}
  T.~G.~Mertens, H.~Verschelde and V.~I.~Zakharov,
  ``Near-Hagedorn Thermodynamics and Random Walks: a General Formalism in Curved Backgrounds,''
  JHEP {\bf 1402}, 127 (2014)
  [arXiv:1305.7443 [hep-th]].

\bibitem{Mertens:2013zya}
  T.~G.~Mertens, H.~Verschelde and V.~I.~Zakharov,
  ``Random Walks in Rindler Spacetime and String Theory at the Tip of the Cigar,''
  JHEP {\bf 1403}, 086 (2014)
  [arXiv:1307.3491 [hep-th]].

\bibitem{Mertens:2014cia}
  T.~G.~Mertens, H.~Verschelde and V.~I.~Zakharov,
  ``Near-Hagedorn Thermodynamics and Random Walks - Extensions and Examples,''
  JHEP {\bf 1411}, 107 (2014)
  [arXiv:1408.6999 [hep-th]].

\bibitem{Mertens:2014dia}
  T.~G.~Mertens, H.~Verschelde and V.~I.~Zakharov,
  ``On the Relevance of the Thermal Scalar,''
  JHEP {\bf 1411}, 157 (2014)
  [arXiv:1408.7012 [hep-th]].


\bibitem{Mertens:2014saa}
  T.~G.~Mertens, H.~Verschelde and V.~I.~Zakharov,
  ``Perturbative String Thermodynamics near Black Hole Horizons,''
  arXiv:1410.8009 [hep-th].

\bibitem{Mertens:2015hia}
  T.~G.~Mertens, H.~Verschelde and V.~I.~Zakharov,
  ``The long string at the stretched horizon and the entropy of large non-extremal black holes,''
  arXiv:1505.04025 [hep-th].

\bibitem{Giribet:2015kca}
  G.~Giribet and A.~Ranjbar,
  ``Screening Stringy Horizons,''
  arXiv:1504.05044 [hep-th].

\bibitem{Dodelson:2015toa}
  M.~Dodelson and E.~Silverstein,
  ``String-theoretic breakdown of effective field theory near black hole horizons,''
  arXiv:1504.05536 [hep-th].


\bibitem{Dodelson:2015uoa}
  M.~Dodelson and E.~Silverstein,
  ``Longitudinal nonlocality in the string S-matrix,''
  arXiv:1504.05537 [hep-th].

\bibitem{Teschner:1999ug}
  J.~Teschner,
  ``Operator product expansion and factorization in the H+(3) WZNW model,''
  Nucl.\ Phys.\ B {\bf 571}, 555 (2000)
  [hep-th/9906215].


\bibitem{Elitzur:1991cb}
  S.~Elitzur, A.~Forge and E.~Rabinovici,
  ``Some global aspects of string compactifications,''
  Nucl.\ Phys.\ B {\bf 359} (1991) 581.


\bibitem{Mandal:1991tz}
  G.~Mandal, A.~M.~Sengupta and S.~R.~Wadia,
  ``Classical solutions of two-dimensional string theory,''
  Mod.\ Phys.\ Lett.\ A {\bf 6} (1991) 1685.

\bibitem{Witten:1991yr}
  E.~Witten,
  ``On string theory and black holes,''
  Phys.\ Rev.\ D {\bf 44} (1991) 314.

\bibitem{Dijkgraaf:1991ba}
  R.~Dijkgraaf, H.~L.~Verlinde and E.~P.~Verlinde,
  ``String propagation in a black hole geometry,''
  Nucl.\ Phys.\ B {\bf 371} (1992) 269.

\bibitem{Bars:1992sr}
  I.~Bars and K.~Sfetsos,
  ``Conformally exact metric and dilaton in string theory on curved space-time,''
  Phys.\ Rev.\ D {\bf 46}, 4510 (1992)
  [hep-th/9206006, hep-th/9206006].

\bibitem{Tseytlin:1993my}
  A.~A.~Tseytlin,
  ``Conformal sigma models corresponding to gauged Wess-Zumino-Witten theories,''
  Nucl.\ Phys.\ B {\bf 411}, 509 (1994)
  [hep-th/9302083].

\bibitem{Aharony:2004xn}
  O.~Aharony, A.~Giveon and D.~Kutasov,
  ``LSZ in LST,''
  Nucl.\ Phys.\ B {\bf 691}, 3 (2004)
  [hep-th/0404016].



\bibitem{Giribet:2000fy}
  G.~Giribet and C.~A.~Nunez,
  ``Aspects of the free field description of string theory on AdS(3),''
  JHEP {\bf 0006}, 033 (2000)
  [hep-th/0006070].

\bibitem{Giribet:2001ft}
  G.~Giribet and C.~A.~Nunez,
  ``Correlators in AdS(3) string theory,''
  JHEP {\bf 0106}, 010 (2001)
  [hep-th/0105200].




\bibitem{Zamolodchikov:1995aa}
  A.~B.~Zamolodchikov and A.~B.~Zamolodchikov,
  ``Structure constants and conformal bootstrap in Liouville field theory,''
  Nucl.\ Phys.\ B {\bf 477}, 577 (1996)
  [hep-th/9506136].

\bibitem{Maldacena:2001kr}
  J.~M.~Maldacena,
  ``Eternal black holes in anti-de Sitter,''
  JHEP {\bf 0304}, 021 (2003)
  [hep-th/0106112].

\bibitem{Barbon:2003aq}
  J.~L.~F.~Barbon and E.~Rabinovici,
  ``Very long time scales and black hole thermal equilibrium,''
  JHEP {\bf 0311}, 047 (2003)
  [hep-th/0308063].

\bibitem{fu} work in progress.

\bibitem{Natsuume:1994sp}
  M.~Natsuume and J.~Polchinski,
  ``Gravitational scattering in the c = 1 matrix model,''
  Nucl.\ Phys.\ B {\bf 424}, 137 (1994)
  [hep-th/9402156].

\bibitem{Gross:1987kza}
  D.~J.~Gross and P.~F.~Mende,
  ``The High-Energy Behavior of String Scattering Amplitudes,''
  Phys.\ Lett.\ B {\bf 197}, 129 (1987).

\bibitem{fzz}
V.A. Fateev, A.B. Zamolodchikov and Al.B. Zamolodchikov, unpublished.

\bibitem{kkk}
V. Kazakov, I. K. Kostov and D. Kutasov, “A Matrix model for the two-dimensional
black hole,” Nucl. Phys. B 622, 141 (2002). [hep-th/0101011].


\bibitem{Itzhaki:2004dv}
  N.~Itzhaki,
  ``The Horizon order parameter,''
  hep-th/0403153.

\bibitem{Giveon:2001up}
  A.~Giveon and D.~Kutasov,
  ``Notes on AdS(3),''
  Nucl.\ Phys.\ B {\bf 621}, 303 (2002)
  [hep-th/0106004].

\bibitem{Bershadsky:1991in}
  M.~Bershadsky and D.~Kutasov,
  ``Comment on gauged WZW theory,''
  Phys.\ Lett.\ B {\bf 266}, 345 (1991).


  \bibitem{Adhikari:1986}
	S. ~K. ~Adhikari,
  ``Quantum Scattering in Two Dimensions,''
  American Journal of Physics {\bf 54}, 362 (1986).

\bibitem{Aharony:1998ub}
  O.~Aharony, M.~Berkooz, D.~Kutasov and N.~Seiberg,
  ``Linear dilatons, NS five-branes and holography,''
  JHEP {\bf 9810}, 004 (1998)
  [hep-th/9808149].

\bibitem{Giveon:1999zm}
  A.~Giveon, D.~Kutasov and O.~Pelc,
  ``Holography for noncritical superstrings,''
  JHEP {\bf 9910}, 035 (1999)
  [hep-th/9907178].

\bibitem{Giveon:1999px}
  A.~Giveon and D.~Kutasov,
  ``Little string theory in a double scaling limit,''
  JHEP {\bf 9910}, 034 (1999)
  [hep-th/9909110].

\bibitem{Giveon:1999tq}
  A.~Giveon and D.~Kutasov,
  ``Comments on double scaled little string theory,''
  JHEP {\bf 0001}, 023 (2000)
  [hep-th/9911039].

\end{thebibliography}
\end{document}